\documentclass[aps,reprint]{revtex4-2}
\usepackage{graphicx}
\usepackage{amsmath}
\usepackage{cases}

\draft % marks overfull lines with a black rule on the right

\begin{document}

%\preprint{}

\title{Statistical regimes of electromagnetic wave propagation in randomly time-varying media}

\author{Seulong Kim}
\affiliation{Research Institute of Basic Sciences, Ajou University, Suwon 16499, South Korea}

\author{Kihong Kim}
\email{khkim@ajou.ac.kr}
\affiliation{Department of Physics, Ajou University, Suwon 16499, South Korea}
\affiliation{School of Physics, Korea Institute for Advanced Study, Seoul 02455, South Korea}
\date{\today}
\begin{abstract}
Wave propagation in time-varying media enables unique control of energy transport by breaking energy conservation through temporal modulation. Among the resulting phenomena, temporal disorder—random fluctuations in material parameters—can suppress propagation and induce localization, analogous to Anderson localization. However, the statistical nature of this process remains incompletely understood.
We present a comprehensive analytical and numerical study of electromagnetic wave propagation in spatially uniform media with randomly time-varying permittivity. Using the invariant imbedding method, we derive exact moment equations and identify three distinct statistical regimes for initially unidirectional input: gamma-distributed energy at early times, negative exponential statistics at intermediate times, and a quasi-log-normal distribution at long times, distinct from the true log-normal. In contrast, symmetric bidirectional input yields genuine log-normal statistics across all time scales.
These findings are validated using two complementary disorder models—delta-correlated Gaussian noise and piecewise-constant fluctuations—demonstrating that the observed statistics are robust and governed by input symmetry. Momentum conservation constrains the long-time behavior, linking the statistical outcome to the initial conditions.
Our results establish a unified framework for understanding statistical wave dynamics in time-modulated systems and offer guiding principles for the design of dynamically tunable photonic and electromagnetic devices.
\end{abstract}

\maketitle

\section{Introduction}

Wave propagation in time-varying media has emerged as a fertile ground for uncovering fundamentally new physical phenomena that are unattainable in static or spatially inhomogeneous systems \cite{shvar,kal,nonreci,gali,most,asga}. In contrast to stationary media, where energy is conserved, time-dependent systems break energy conservation due to the explicit temporal variation of material parameters. This leads to unconventional scattering dynamics, temporal energy amplification, and novel mechanisms for controlling wave transport. Such effects are increasingly relevant in modern optics, materials science, and communication technologies, where temporal modulation enables dynamic functionalities beyond the capabilities of static systems.

Recent advances have revealed a variety of phenomena unique to time-varying systems, including temporal photonic crystals, momentum band gaps, temporal Brewster effects, temporal holography, and ultrafast wave steering \cite{zur,vald,lust,tv8,tv9,tv10,pac2,solis,kouf,mir,kgap,dr1,sk1,tv2,miya,tv6}. Among these, temporal disorder—random fluctuations in material properties over time—has drawn particular attention for its striking analogy to Anderson localization in spatially disordered media \cite{shar,carm,apf,garn,jkim,bo,eswa,kim2}. Temporal disorder can induce strong localization effects, leading to suppression of wave propagation. These effects are of both fundamental interest and practical importance, with potential applications in ultrafast switching, energy harvesting, and temporally gated information processing.

Despite growing interest, the statistical behavior of waves in randomly time-varying media remains incompletely understood. Previous studies suggest that energy distributions may evolve from negative exponential forms at short times to log-normal distributions at longer times \cite{carm}. However, key questions remain unresolved: What governs the transition between these regimes? Is log-normal behavior truly universal? And how do initial wave conditions influence the statistical evolution?

In this work, we present a comprehensive analytical and numerical study of electromagnetic wave propagation in isotropic, spatially uniform media with randomly time-varying permittivity. Using the invariant imbedding method (IIM) \cite{kim2,bell,gol,ramm,ramm2,klya,kk0,sk2,sk5}, we derive exact moment equations for reflectance, transmittance, and total energy density. For unidirectional input, our analysis reveals three distinct statistical regimes: gamma-distributed energy at early times, negative exponential behavior at intermediate times, and a quasi-log-normal distribution at long times, which is mathematically distinct from the true log-normal. In contrast, symmetric bidirectional input produces genuine log-normal statistics across all time scales.

These findings are supported by extensive calculations based on two complementary disorder models: delta-correlated Gaussian noise and piecewise-constant stepwise fluctuations. Both models confirm that wave statistics are highly sensitive to the initial input symmetry and the duration of temporal disorder. Our results establish a unified framework for understanding statistical fluctuations and localization phenomena in time-varying systems and provide practical guidance for the design of dynamically tunable photonic and electromagnetic devices.

\section{Theory}

We investigate electromagnetic wave propagation in an isotropic, spatially uniform medium with time-dependent properties. The electric displacement field
$D$ satisfies the wave equation
\begin{align}
  \frac{d}{dt}\left(\mu \frac{dD}{dt}\right)+\frac{\omega_0^2}{\epsilon}D=0,~\omega_0=ck,
\end{align}
where
$\epsilon$ and $\mu$ are the permittivity and permeability, respectively, $c$ is the speed of light in vacuum, and
$k$ is the constant wavenumber. We consider wave propagation along the
$x$ axis.

In a time-independent medium, the field components evolve as $\exp[i(kx \mp \omega t)]$. When material parameters vary with time, scattering occurs at temporal interfaces, generating reflected and transmitted components. We consider a unit-amplitude plane wave initially propagating in the $+x$ direction. If $\epsilon$ and $\mu$ vary only within the interval $0 \le t \le T$, the displacement field is given by
\begin{equation}
D(t) =
\begin{cases}
e^{-i\omega_1 t}, & t < 0 \\
s(T)e^{-i\omega_2 (t - T)} + r(T)e^{i\omega_2 (t - T)}, & t > T
\end{cases},
\label{eq:iniw}
\end{equation}
where $\omega_1 = \omega_0 / \sqrt{\epsilon_1 \mu_1}$ and $\omega_2 = \omega_0 / \sqrt{\epsilon_2 \mu_2}$, with $\epsilon_{1,2}$ and $\mu_{1,2}$ denoting the permittivity and permeability before and after the modulation, respectively. All parameters are assumed to be positive. The coefficients $s(T)$ and $r(T)$ represent the temporal transmission and reflection amplitudes.

We analyze this system using the IIM \cite{kim2,bell,gol,ramm,ramm2,klya,kk0,sk2,sk5}, a powerful tool originally developed for spatially inhomogeneous systems. In the IIM, the independent variable is not time
$t$ itself but the modulation interval
$T$. The method transforms the boundary value problem into an initial value problem, yielding coupled differential equations for
$r$ and
$s$ with respect to the imbedding parameter
$\tau\in [0,T]$, which serves as the running duration of temporal variation:
\begin{equation}
   \frac{1}{\omega_0}\frac{dr}{d\tau}=i\beta r+i\alpha s,~
   \frac{1}{\omega_0}\frac{ds}{d\tau}=-i\alpha r-i\beta s,
   \label{eq:iie}
   \end{equation}
   where
 \begin{equation}
   \alpha=\frac{1}{2n_2}\left(\frac{\epsilon_2}{\epsilon}-\frac{\mu_2}{\mu}\right),
   ~\beta=\frac{1}{2n_2}\left(\frac{\epsilon_2}{\epsilon}+\frac{\mu_2}{\mu}\right),
   \label{eq:ab}
 \end{equation}
and
$n_2=\sqrt{\epsilon_2\mu_2}$. The initial conditions at
$\tau=0$, corresponding to a sudden temporal interface, are given by
\begin{align}
   r(0)=\frac{1}{2}\left(1-\sqrt{\frac{\epsilon_2\mu_1}{\epsilon_1\mu_2}}\right),~
   s(0)=\frac{1}{2}\left(1+\sqrt{\frac{\epsilon_2\mu_1}{\epsilon_1\mu_2}}\right).
   \label{eq:ic}
 \end{align}

From Eq.~\eqref{eq:iie}, it follows that the difference between the squared magnitudes of the transmission and reflection amplitudes remains invariant during the temporal modulation:
\begin{align}
   \vert s(\tau)\vert^2-\vert r(\tau)\vert^2=\vert s(0)\vert^2-\vert r(0)\vert^2=\sqrt{\frac{\epsilon_2\mu_1}{\epsilon_1\mu_2}}.
 \end{align}
We define the temporal transmittance
$S$ and reflectance
$R$ as
\begin{align}
   S=\sqrt{\frac{\epsilon_1\mu_2}{\epsilon_2\mu_1}}\vert s\vert^2,~
   R=\sqrt{\frac{\epsilon_1\mu_2}{\epsilon_2\mu_1}}\vert r\vert^2.
   \label{eq:defrs}
 \end{align}
It immediately follows that $S - R = 1$ for arbitrary temporal variations of $\epsilon$ and $\mu$. Since both $S$ and $R$ are nonnegative, we always have $S \geq 1$, indicating that total temporal reflection is not possible in isotropic media. In the special case of perfect transmission ($R=0$), we find $S = 1$, implying no amplification.

In Eq.~(\ref{eq:iniw}), $s$ and $r$ are defined relative to the displacement field $D$ of the initial wave. The energy density is $\vert D\vert^2/(2\epsilon)$ and the photon energy is $\hbar\omega=\hbar ck/\sqrt{\epsilon\mu}$, giving a photon number density
\begin{align}
    \frac{\vert D\vert^2}{2\epsilon}\cdot \frac{\sqrt{\epsilon\mu}}{\hbar ck}=\sqrt{\frac{\mu}{\epsilon}}\frac{\vert D\vert^2}{2\hbar ck}.
\end{align}
Thus, the definitions of $S$ and $R$ in Eq.~(\ref{eq:defrs})
correspond to the ratios of photon number densities in the transmitted and reflected waves to that of the initial wave.

After temporal scattering, equal numbers of forward- and backward-moving photons are created or annihilated. The forward photon carries momentum $+\hbar k$ and the backward photon $-\hbar k$, so the total momentum density relative to the initial wave is $\hbar k(S-R)$, which remains constant. The relation $S-R=1$ thus directly reflects momentum conservation.

Alternatively, transmittance and reflectance may be defined in terms of the energy flux, given by the product of the energy density and the wave velocity:
\begin{align}
    S_E=\frac{\epsilon_1\sqrt{\epsilon_1\mu_1}}{\epsilon_2\sqrt{\epsilon_2\mu_2}}\vert s\vert^2,~R_E=\frac{\epsilon_1\sqrt{\epsilon_1\mu_1}}{\epsilon_2\sqrt{\epsilon_2\mu_2}}\vert r\vert^2.
\end{align}
In this formulation, $S_E-R_E$ is generally not conserved, except in the special case $\epsilon_1\mu_1=\epsilon_2\mu_2$, for which $S=S_E$ and $R=R_E$.

In this study, we examine electromagnetic wave propagation in media where the dielectric permittivity $\epsilon(t)$ varies randomly in time. Although our method can be readily extended to cases where both
$\epsilon$ and
$\mu$ fluctuate stochastically, we fix
$\mu=1$ throughout this work for simplicity. Two stochastic models are considered.
Model 1 assumes delta-correlated Gaussian noise:
\begin{align}
   &\epsilon(t)=\bar{\epsilon}+\delta \epsilon(t),\nonumber\\  &\langle\delta\epsilon(t)\delta\epsilon(t^\prime)\rangle=g_0\delta(t-t^\prime),
   ~\langle\delta\epsilon(t)\rangle=0,
  \end{align}
where $\bar{\epsilon}$ is the mean permittivity and $g_0$ (with units of time) quantifies the disorder strength. This model is analytically tractable and enables semi-analytical evaluation of all disorder-averaged moments. Although
$\epsilon(t)$ may temporarily take negative values in this model, it does not produce any distinctive effects in the weak-disorder cases considered in the present work. Model 2 describes piecewise-constant disorder: the fluctuation $\delta\epsilon$ is constant within each time interval of duration $\Lambda$, and is abruptly updated at the end of each interval. Each new value is drawn independently from a uniform distribution over $[-a_0,a_0]$. The process continues over a total duration $T$. This model is particularly well suited for computing statistical quantities that are difficult to evaluate using Model 1, such as the moments of the logarithm of the wave energy and the full probability distributions of the logarithms of both the reflectance and the wave energy.

In Model 1, we study the stochastic differential equation [Eq.~(\ref{eq:iie})] with random coefficients proportional to
$1/\epsilon$. For a random function in the denominator, we assume weak disorder and expand the inverse permittivity as
\begin{align}
   \frac{1}{\epsilon}=\frac{1}{\bar{\epsilon}+\delta\epsilon}
   \approx\frac{1}{\bar{\epsilon}}-\frac{\delta \epsilon}{\bar{\epsilon}^2}.
   \label{eq:approx}
  \end{align}
Alternatively, $\epsilon$ may be modeled as a dichotomous random variable; in this case, even for strong disorder, the disorder average can be performed exactly using the Shapiro–Loginov formula of differentiation \cite{shalog}.

Our goal is to compute the statistical moments of the transmittance, reflectance, and total wave energy density. To this end, we define the moment function
$Z_{abcd}=\left\langle r^a{\left(r^*\right)}^bs^c{\left(s^*\right)}^d\right\rangle$, where $a,b,c,d=0,1,2,\cdots$.
Applying Eq.~\eqref{eq:iie} together with the Furutsu–Novikov formula \cite{fur,nov}, we derive the following evolution equation for $Z_{abcd}$:
\begin{align}
   &\frac{1}{\omega_0}\frac{d}{d\tau}Z_{abcd}=C_1Z_{abcd}+C_2Z_{a+1,b,c-1,d}\nonumber\\
   &~~~+C_3Z_{a-1,b,c+1,d}+C_4Z_{a,b+1,c,d-1}\nonumber\\
   &~~~+C_5Z_{a,b-1,c,d+1}+C_6Z_{a+1,b+1,c-1,d-1}\nonumber\\
   &~~~+C_7Z_{a-1,b-1,c+1,d+1}+C_8Z_{a+1,b-1,c-1,d+1}\nonumber\\
   &~~~+C_9Z_{a-1,b+1,c+1,d-1}+C_{10}Z_{a+2,b,c-2,d}\nonumber\\
   &~~~+C_{11}Z_{a-2,b,c+2,d}+C_{12}Z_{a,b+2,c,d-2}\nonumber\\
   &~~~+C_{13}Z_{a,b-2,c,d+2},
   \label{eq:z}
  \end{align}
where the coefficients are given by
  \begin{align}
  & C_1=i\bar{\beta}\left(a-b-c+d\right)+\frac{g\gamma^2}{2}\Big[a+b+c+d\nonumber\\
  &~~~~~~~ +2ac+2bd-\left(a-b-c+d\right)^2\Big],\nonumber\\
  & C_2=c\left[-i\bar{\alpha}+g\gamma^2\left(a-b-c+d+1\right)\right],\nonumber\\
  & C_3=a\left[i\bar{\alpha}-g\gamma^2\left(a-b-c+d-1\right)\right],\nonumber\\
  & C_4=d\left[i\bar{\alpha}-g\gamma^2\left(a-b-c+d-1\right)\right],\nonumber\\
  & C_5=b\left[-i\bar{\alpha}+g\gamma^2\left(a-b-c+d+1\right)\right],\nonumber\\
  & C_6=g\gamma^2cd,~C_7=g\gamma^2ab,\nonumber\\ & C_8=-g\gamma^2bc,~C_9=-g\gamma^2ad,\nonumber\\
  & C_{10}=-\frac{g\gamma^2}{2}c\left(c-1\right),~C_{11}=-\frac{g\gamma^2}{2}a\left(a-1\right),\nonumber\\
  & C_{12}=-\frac{g\gamma^2}{2}d\left(d-1\right),~C_{13}=-\frac{g\gamma^2}{2}b\left(b-1\right).
  \end{align}
The auxiliary parameters are defined as
   \begin{align}
   &\bar{\alpha}=\frac{1}{2n_2}\left(\frac{\epsilon_2}{\bar{\epsilon}}-\frac{\mu_2}{\mu}\right),~
   \bar{\beta}=\frac{1}{2n_2}\left(\frac{\epsilon_2}{\bar{\epsilon}}+\frac{\mu_2}{\mu}\right),\nonumber\\
   &\gamma=\frac{1}{2n_2}\frac{\epsilon_2}{\bar{\epsilon}^2},~g=g_0\omega_0.
   \label{eq:param}
  \end{align}

The IIM can be extended to more general cases in which plane waves with arbitrary relative amplitudes propagate simultaneously in opposite directions at the initial time. In this case, we generalize Eq.~(\ref{eq:iniw}) and the initial conditions in Eq.~(\ref{eq:ic}) as follows:
 \begin{align}
 &D=\begin{cases}
         {v}e^{-i\omega_1t}+{w}e^{i\omega_1t}, & \mbox{if } t<0 \\
        s(T)e^{-i\omega_2 (t-T)}+r(T)e^{i\omega_2 (t-T)}, & \mbox{if } t>T
       \end{cases},\nonumber\\
       &r(0)=\frac{{v}}{2}\left(1-\sqrt{\frac{\epsilon_2\mu_1}{\epsilon_1\mu_2}}\right)
       +\frac{{w}}{2}\left(1+\sqrt{\frac{\epsilon_2\mu_1}{\epsilon_1\mu_2}}\right),\nonumber\\
       &s(0)=\frac{{v}}{2}\left(1+\sqrt{\frac{\epsilon_2\mu_1}{\epsilon_1\mu_2}}\right)
       +\frac{{w}}{2}\left(1-\sqrt{\frac{\epsilon_2\mu_1}{\epsilon_1\mu_2}}\right),
       \label{eq:gic}
   \end{align}
where ${v}$ and ${w}$
denote the amplitudes of the incident waves propagating in the $+x$ and $-x$ directions, respectively, for $t<0$.
Under these generalized initial conditions, the invariant imbedding equations in Eqs.~(\ref{eq:iie}) and (\ref{eq:z}) remain valid and unchanged.

The IIM can be viewed as a continuum generalization of the transfer matrix method widely used in wave propagation studies and offers several advantages over alternative approaches. First, for temporal variations described by continuous or piecewise continuous functions, the invariant imbedding equations can be integrated directly using standard differential equation solvers, eliminating the need for artificial discretization. Second, for random temporal variations, the method can incorporate stochastic calculus tools—such as the Furutsu–Novikov formula or the Shapiro–Loginov formula of differentiation—to perform analytical disorder averaging, as shown in the derivation of Eq.~(\ref{eq:z}). This transforms the original equations with random coefficients into a larger set of coupled equations with deterministic coefficients, enabling direct solutions without numerical averaging over many realizations. This often results in substantial computational efficiency and, in some cases, allows for closed-form analytical solutions \cite{sk5}. Third, in nonlinear systems, the formalism introduces an additional invariant imbedding equation for the wave intensity, which naturally captures nonlinear effects such as bistability and multistability and enhances the efficiency of both analytical and numerical treatments \cite{ramm2,klya,kk0}.

\begin{figure}
    \centering
    \includegraphics[width=\columnwidth]{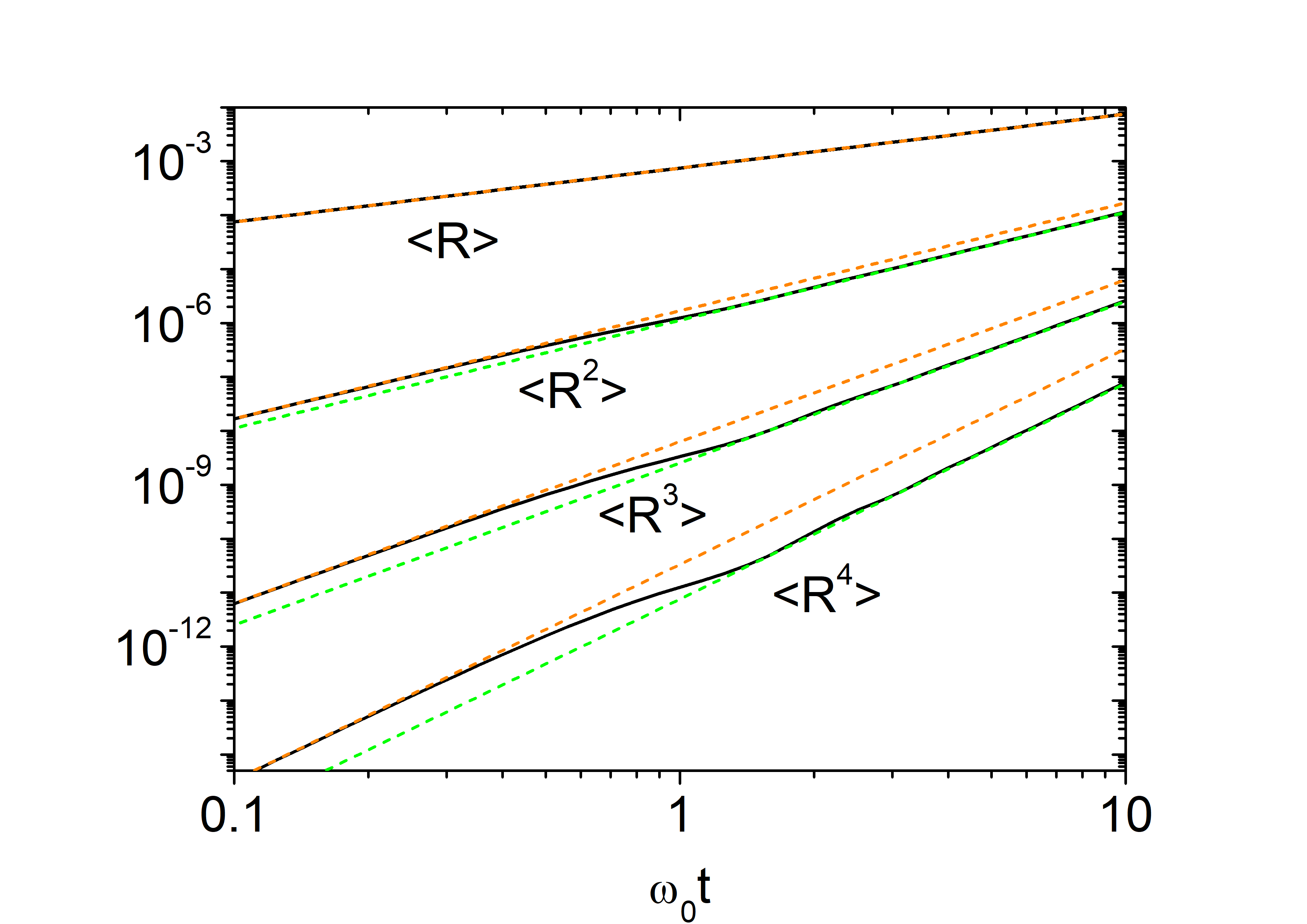}
    \caption{Log-log plot of the temporal evolution of the reflectance moments $\langle R \rangle$, $\langle R^2 \rangle$, $\langle R^3 \rangle$, and $\langle R^4 \rangle$ for Model 1 with $g = 0.003$. Numerical results from Eq.~(\ref{eq:z}) are shown as black solid lines. The orange dashed lines represent the short-time behavior predicted by Eq.~(\ref{eq:short}), while the green dashed lines correspond to the intermediate-time behavior described by Eq.~(\ref{eq:exp}). For $n\ge 2$, a clear crossover from the initial regime governed by Eq.~(\ref{eq:short}) to the regime of Eq.~(\ref{eq:exp}) occurs near $\omega_0t\approx 1$.
    }\label{f2}
  \end{figure}

\section{Results}

We present the results of our analytical and numerical calculations. For simplicity, we focus on Model 1 with
$\epsilon_1 = \epsilon_2 = \bar{\epsilon} = 1$ and $\mu_1 = \mu_2=\mu=1$, under which the parameters in Eq.~(\ref{eq:param}) simplify to $\bar{\alpha}=0$, $\bar{\beta}=1$, and $\gamma=1/2$. In this case, the reflectance and transmittance reduce to $R=\vert r\vert^2$ and $S=\vert s\vert^2$, respectively.

The structure of Eq.~(\ref{eq:z}) couples each moment $Z_{abcd}$ to others with the same total indices, satisfying $a^\prime+c^\prime=a+c$ and $b^\prime+d^\prime=b+d$. To compute $Z_{nn00}=\langle R^{n}\rangle$ and $Z_{00nn}=\langle S^{n}\rangle$, it suffices to solve a finite system of $(n + 1)^2$ coupled first-order differential equations for the moments $Z_{i,j,n-i,n-j}$, where $i,j=0,1,\cdots,n$. Since the coefficients in these equations are constant, they are, in principle, analytically solvable; however, explicit solutions are generally intractable. We therefore obtain numerically precise values of $\langle R^{n}\rangle$ and $\langle S^{n}\rangle$ by solving the $(n + 1)^2$ coupled equations derived from Eq.~(\ref{eq:z}) using a Fortran-based code combined with IMSL library routines developed by the authors.

When $\epsilon_1=\epsilon_2$ and $\mu_1=\mu_2$, the quantities $R$ and $S$ represent not only the photon number densities but also the reflected and transmitted energy densities, respectively, normalized to that of the initial wave. As seen from Eq.~(\ref{eq:iniw}), the total energy density relative to the initial wave for $t>T$, denoted $U$, is given by $U=R+S=2R+1$. This expression follows from the standard practice in electromagnetic wave propagation of defining both energy density and energy flux as time-averaged quantities, particularly for monochromatic waves. The temporal evolution of statistical moments depends on the time regime—short, intermediate, or long. For studying early-time dynamics, reflectance is a more sensitive observable than total energy, since $R(0)=0$ while $U(0)=1$.

The evolution also depends on the initial wave conditions.
In Fig.~\ref{f2}, we plot the first four moments of the reflectance in the short-time regime, assuming the wave initially propagates only in the forward direction [i.e., $r(0)={w}=0$, $s(0)={v}=1$].
The corresponding initial condition is $Z_{abcd}(0)=1$ if $a=b=0$, and zero otherwise.

We find that, at early times, the reflectance moments are
\begin{align}
   \left\langle R^n\right\rangle=\left(2n-1\right)!!\left(\frac{g\omega_0t}{4}\right)^n,
   \label{eq:short}
  \end{align}
while for $\omega_0t\approx 1$ they cross over to
 \begin{align}
   \left\langle R^n\right\rangle=n!\left(\frac{g\omega_0t}{4}\right)^n,
   \label{eq:exp}
  \end{align}
signaling a change in the underlying probability distribution. These behaviors are confirmed numerically and derived analytically in the Appendix. At early times, $R$ follows a gamma distribution:
\begin{equation}
P(R)=\frac{1}{\sqrt{\pi\theta R} }e^{-R/\theta},\quad \theta=\frac{g\omega_0t}{2},
\end{equation}
with moments
\begin{align}
\langle R^n\rangle=\theta^n\frac{\Gamma\left(n+\frac{1}{2}\right)}{\Gamma\left(\frac{1}{2}\right)}.
  \end{align}
As $\omega_0t$ grows to order unity, this distribution transitions to a negative exponential form,
\begin{equation}
    P(R)=\frac{2}{\theta}{e^{-2R/\theta}},
\end{equation}
whose moments are given by Eq.~(\ref{eq:exp}). This regime persists up to $ \omega_0 t \lesssim g^{-1}$.

Under the random phase approximation (RPA), which assumes
\begin{align}
   Z_{abcd}=0,\text{~when } a\ne b\text{ and }c\ne d,
   \label{eq:rpa}
  \end{align}
the reflectance moments exactly match those of the  negative exponential distribution. This result, consistent with \cite{carm}, confirms that under RPA, the reflectance indeed follows a negative exponential distribution.

In summary, in the intermediate regime $1 \lesssim \omega_0 t \lesssim g^{-1}$, the reflectance moments converge to those of a negative exponential distribution. The transition from gamma to negative exponential, occurring near $\omega_0 t \sim 1$, is driven by complete randomization of the phases of $r$ and $s$, and is largely insensitive to disorder strength in the weak-disorder regime. The crossover time is inversely proportional to $\omega_0 = ck$.

  \begin{figure}
    \centering
    \includegraphics[width=\columnwidth]{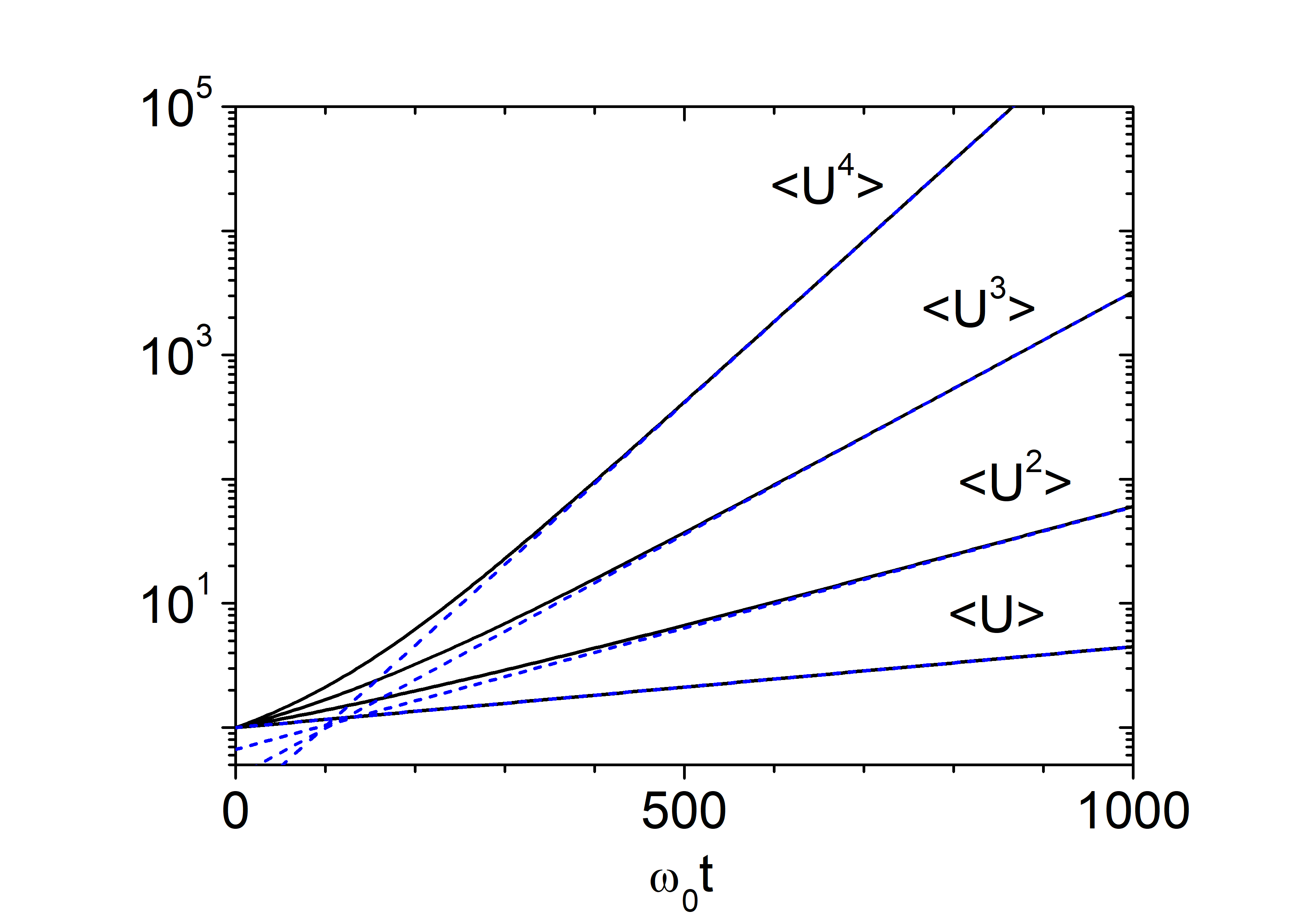}
    \caption{Temporal growth of energy moments for $g = 0.003$.
The evolution of $\langle U \rangle$, $\langle U^2 \rangle$, $\langle U^3 \rangle$, and $\langle U^4 \rangle$ is shown. Solid lines represent numerical results from Eq.~(\ref{eq:z}), and dashed lines show the analytical prediction from Eq.~(\ref{eq:u1}). Excellent agreement is observed in the long-time regime $\omega_0 t \gtrsim g^{-1}$.
}
 \label{f3}
  \end{figure}

We now examine the time evolution of the total wave energy, focusing on the long-time regime.
In Fig.~\ref{f3}, we present the first four moments of $U$, which show excellent agreement with the analytical expression
\begin{align}
   \left\langle U^n\right\rangle=\frac{n!}{\left(2n-1\right)!!}\exp{\left[n(n+1)\left(\frac{g\omega_0t}{4}\right)\right]},
   \label{eq:u1}
  \end{align}
  valid over a broad temporal range,
$g^{-1} \lesssim\omega_0 t\lesssim 10^6$.
  The case $n=1$ is special. Numerical results confirm that the average energy is accurately described by
\begin{align}
   \left\langle U\right\rangle=\exp{\left(\frac{g\omega_0t}{2}\right)},
   \label{eq:energy}
  \end{align}
throughout the entire interval $0<\omega_0t\lesssim 10^6$.

Although Eq.~(\ref{eq:u1}) has the same exponential factor as the raw moments of a log-normal distribution,
\begin{align}
   \left\langle U^n\right\rangle_{\rm LN}=\exp{\left[n(n+1)\left(\frac{g\omega_0t}{4}\right)\right]},
   \label{eq:u2}
  \end{align}
the additional prefactor $n!/(2n-1)!!$ in Eq.~(\ref{eq:u1})
distinguishes the actual distribution from a true log-normal form. This prefactor modifies all moments, altering the peak position, width, and tail behavior of the probability density. While the two distributions become asymptotically similar as $t\rightarrow\infty$, substantial and experimentally relevant deviations persist over a wide time range. We therefore classify the underlying statistics not as strictly log-normal, but as a quasi-log-normal distribution.

The exponential factor common to both forms can be obtained in a semi-analytical manner from Eq.~(\ref{eq:z}).
From full numerical solutions of Eq.~(\ref{eq:z}), we find that, in calculating $Z_{nn00}=\langle R^n\rangle$, all terms with $a\ne b$ or $c\ne d$ in $Z_{abcd}$ decay rapidly at long times. This yields
\begin{align}
    \frac{1}{\omega_0}\frac{d}{d\tau}\langle R^n\rangle&\approx g\gamma^2 n\langle R^n\rangle + g\gamma^2 n^2 \langle R^{n-1}S\rangle
    \nonumber\\&=\frac{g}{4} n(n+1)\langle R^n\rangle
    +\frac{g}{4} n^2 \langle R^{n-1}\rangle,
\end{align}
where $S=R+1$ and $\gamma=1/2$. If we further incorporate the numerical observation that $\langle R^n\rangle$ increases rapidly with $n$ in the long-time regime,
we obtain
\begin{align}
   \langle R^n\rangle,~\langle S^n\rangle,~\langle U^n\rangle\propto \exp{\left[n(n+1)\left(\frac{g\omega_0t}{4}\right)\right]},
   \label{eq:uscal}
\end{align}
with $\tau$ replaced by $t$.
This reasoning establishes the exponential scaling but not the numerical prefactor, which is governed by the evolution of the rapidly decaying terms in Eq.~(\ref{eq:z}) and further constrained by the
$w/v$ ratio in Eq.~(\ref{eq:gic}), as discussed below.

    \begin{figure}
    \centering
    \includegraphics[width=\columnwidth]{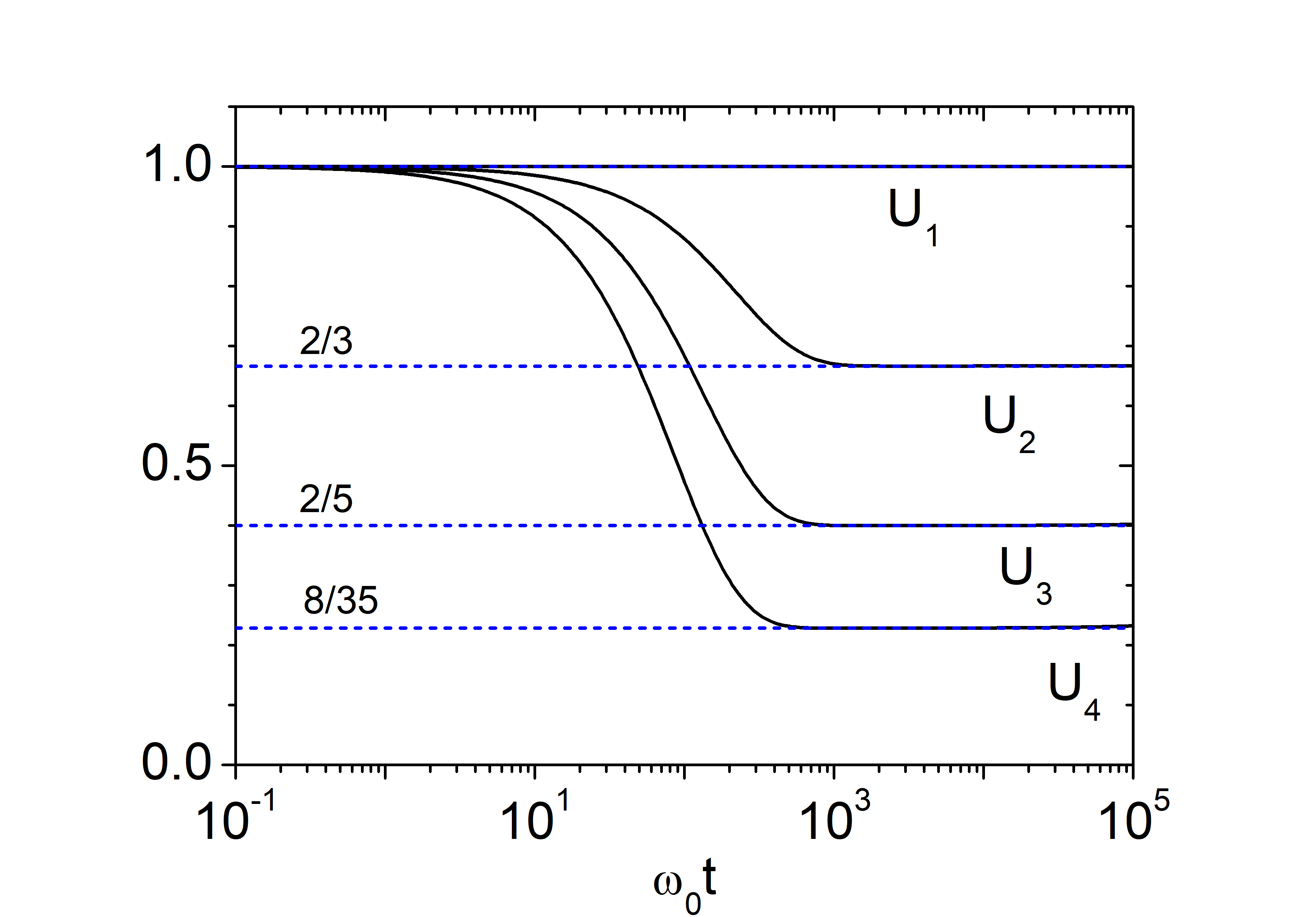}
    \caption{Temporal variation of $U_n$ for $g = 0.003$.
Here, $U_n = \langle U^n \rangle \exp\left[-n(n+1)g\omega_0 t / 4\right]$. In the long-time regime, numerical results show excellent agreement with the analytical predictions from Eq.~(\ref{eq:u1}).}
    \label{ff3}
  \end{figure}

 \begin{figure}
    \centering
    \includegraphics[width=\columnwidth]{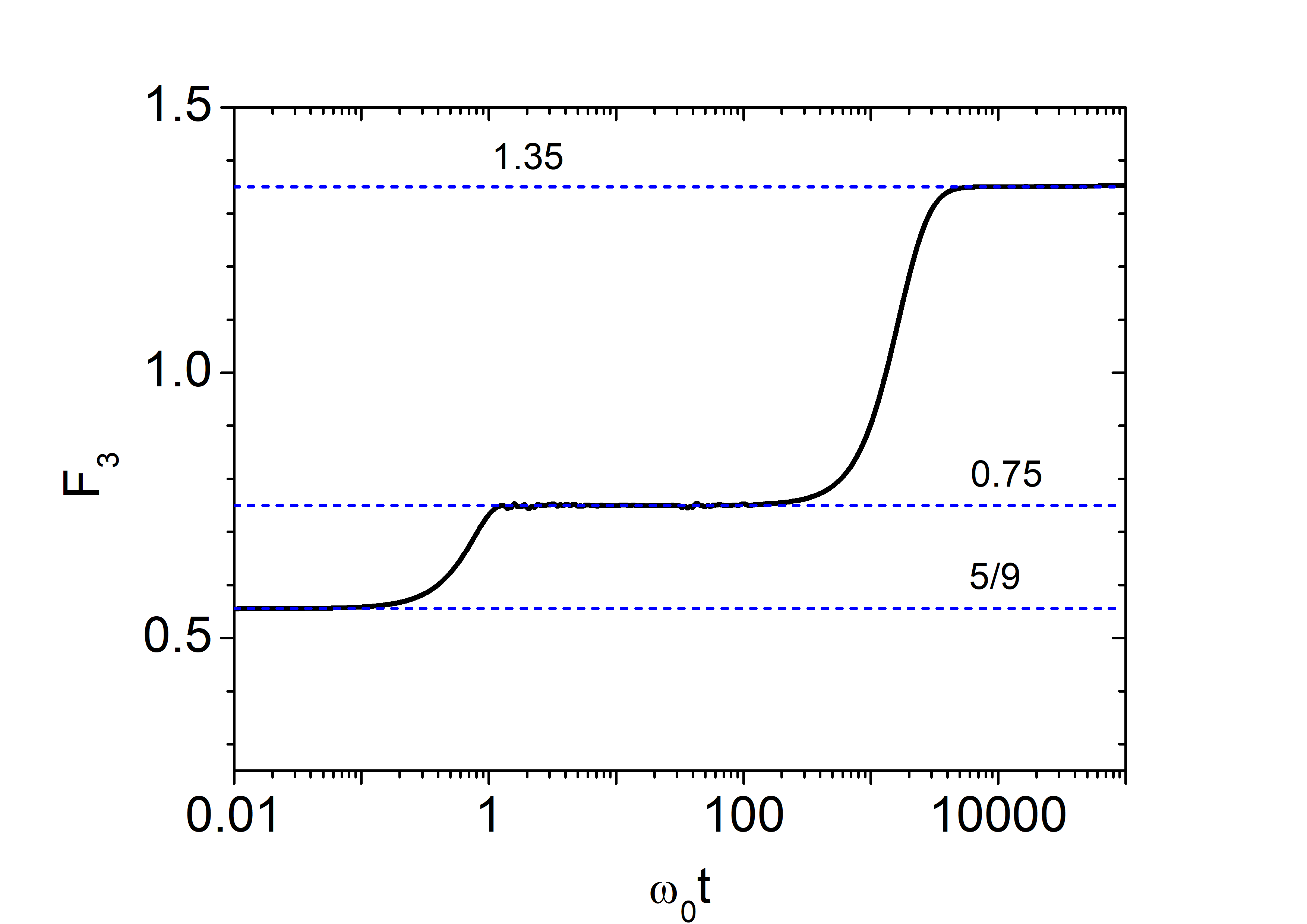}
    \caption{Temporal evolution of $F_3$ for $g = 0.003$.
$F_3 \equiv \langle R^3 \rangle \langle R \rangle^3 / \langle R^2 \rangle^3$ evolves from an initial value of 5/9 (gamma distribution), passes through 0.75 (negative exponential
distribution), and approaches 1.35, consistent with the quasi-log-normal form given in Eq.~(\ref{eq:u1}).}
    \label{ff4}
  \end{figure}

  \begin{figure}
    \centering
    \includegraphics[width=\columnwidth]{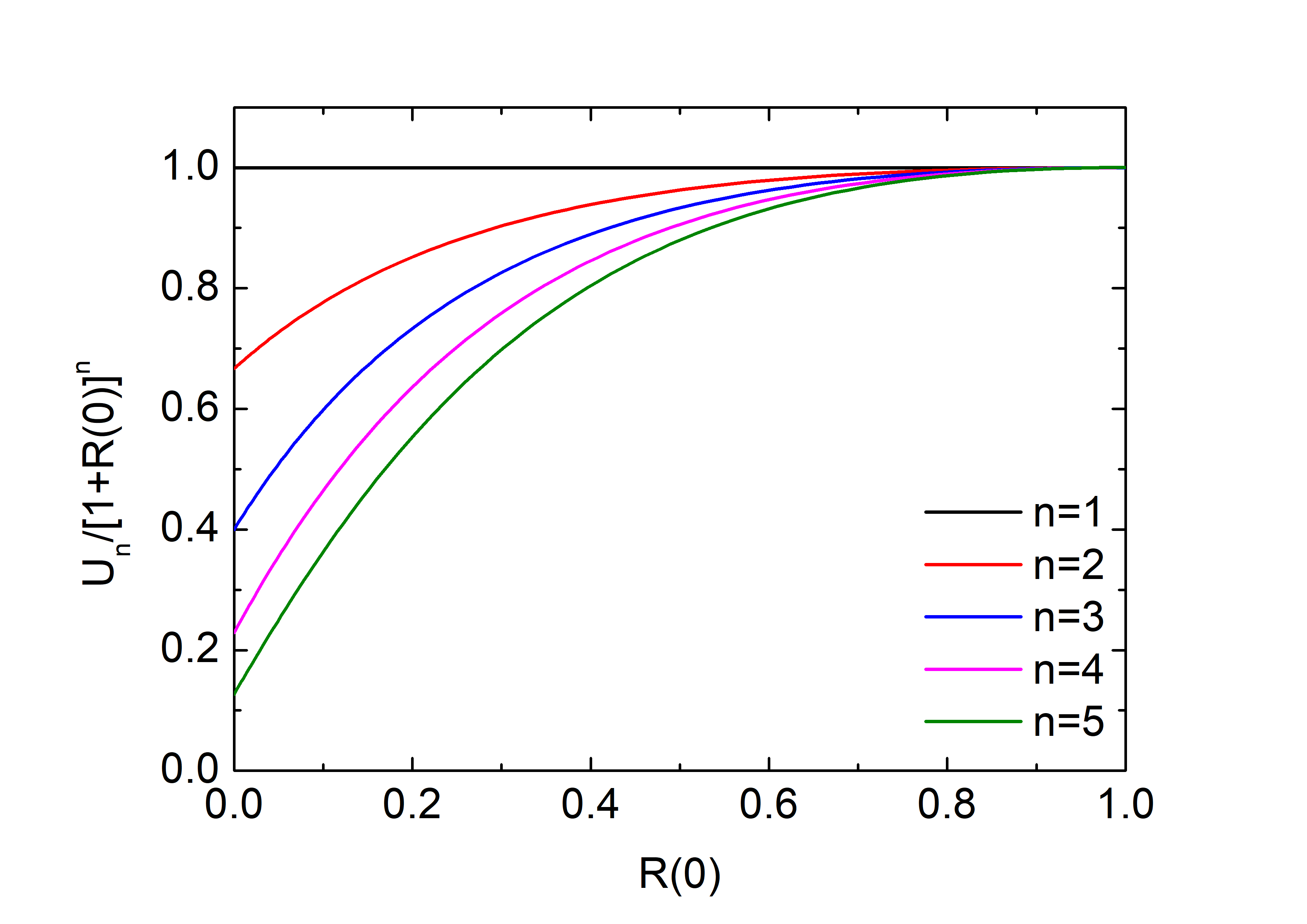}
    \caption{Scaled moments $U_n / [1 + R(0)]^n$ as functions of $R(0) = |{w}|^2$ for $n = 1$–5 at $\omega_0 t = 25000$. $S(0) = |{v}|^2$ is fixed at 1.
    }
    \label{ff5}
  \end{figure}

   \begin{figure}
    \centering
    \includegraphics[width=\columnwidth]{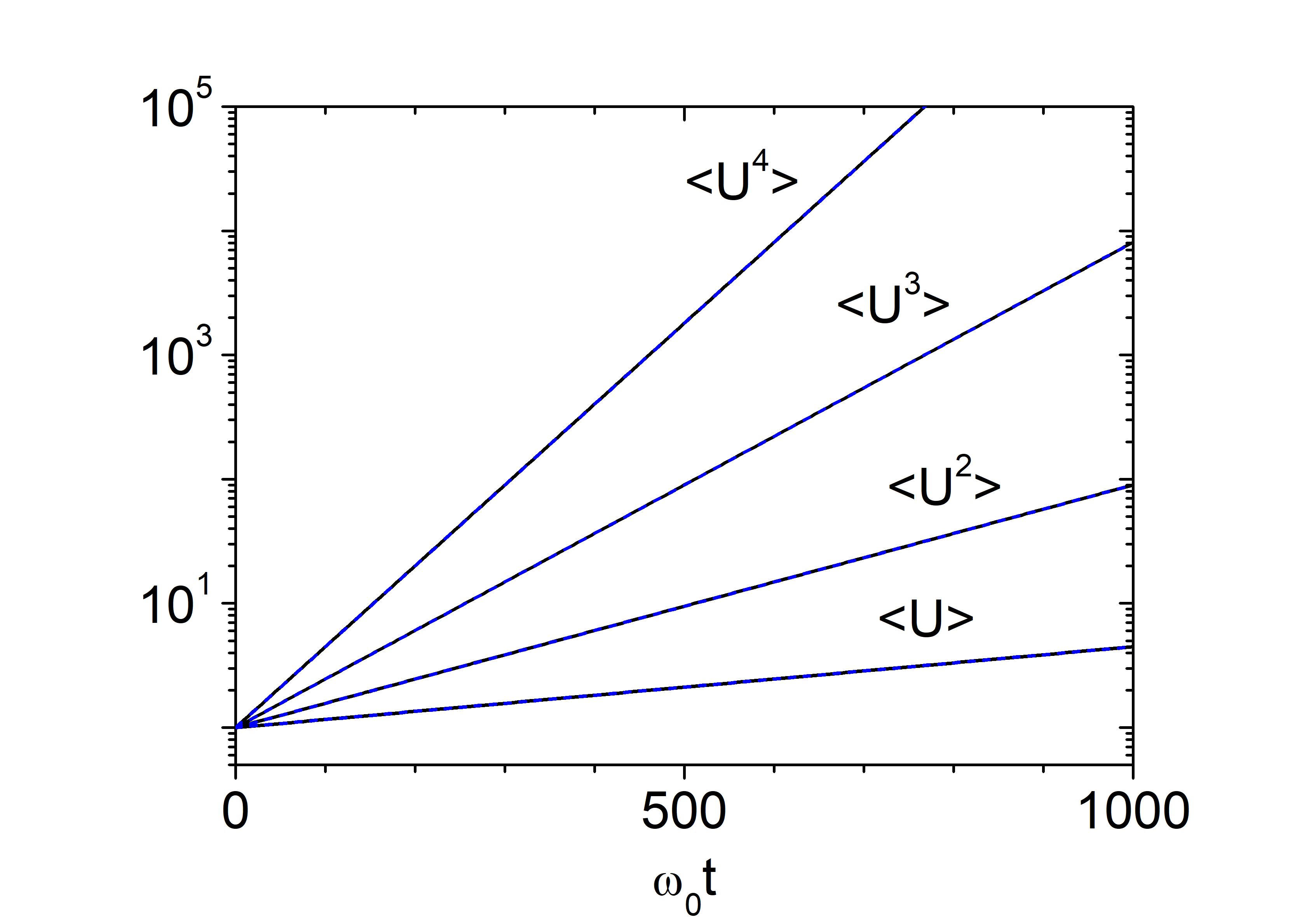}
    \caption{Temporal growth of energy moments for $g = 0.003$ with $R(0) = S(0) = 1/2$.
Solid lines show numerical results from Eq.~(\ref{eq:z}); dashed lines indicate the analytical prediction from Eq~(\ref{eq:u2}). The results confirm that the energy follows a log-normal distribution at all times when the initial waves propagate in opposite directions with equal amplitudes.}
 \label{ff6}
  \end{figure}
  
    \begin{figure}
    \centering
    \includegraphics[width=\columnwidth]{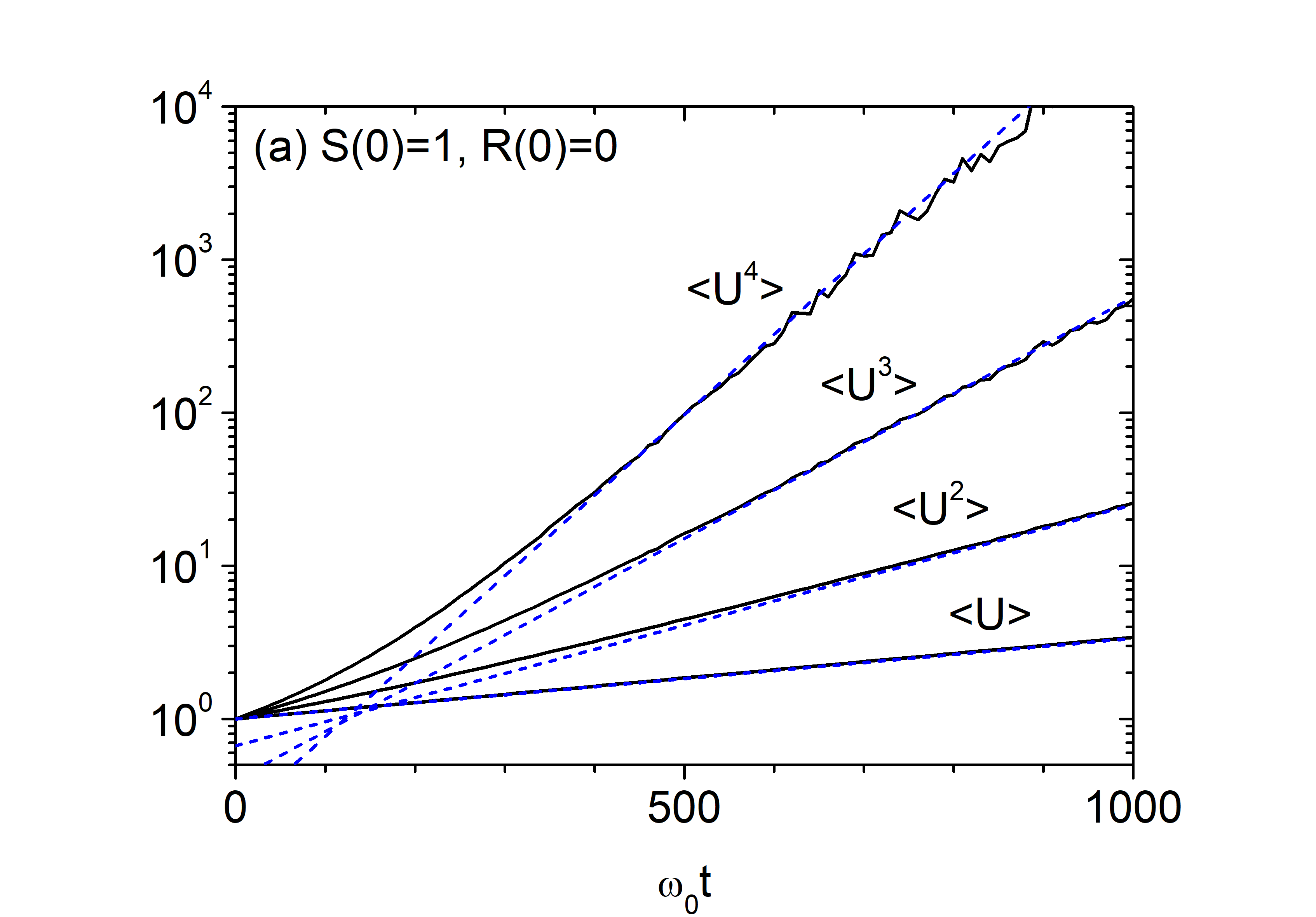}
   \includegraphics[width=\columnwidth]{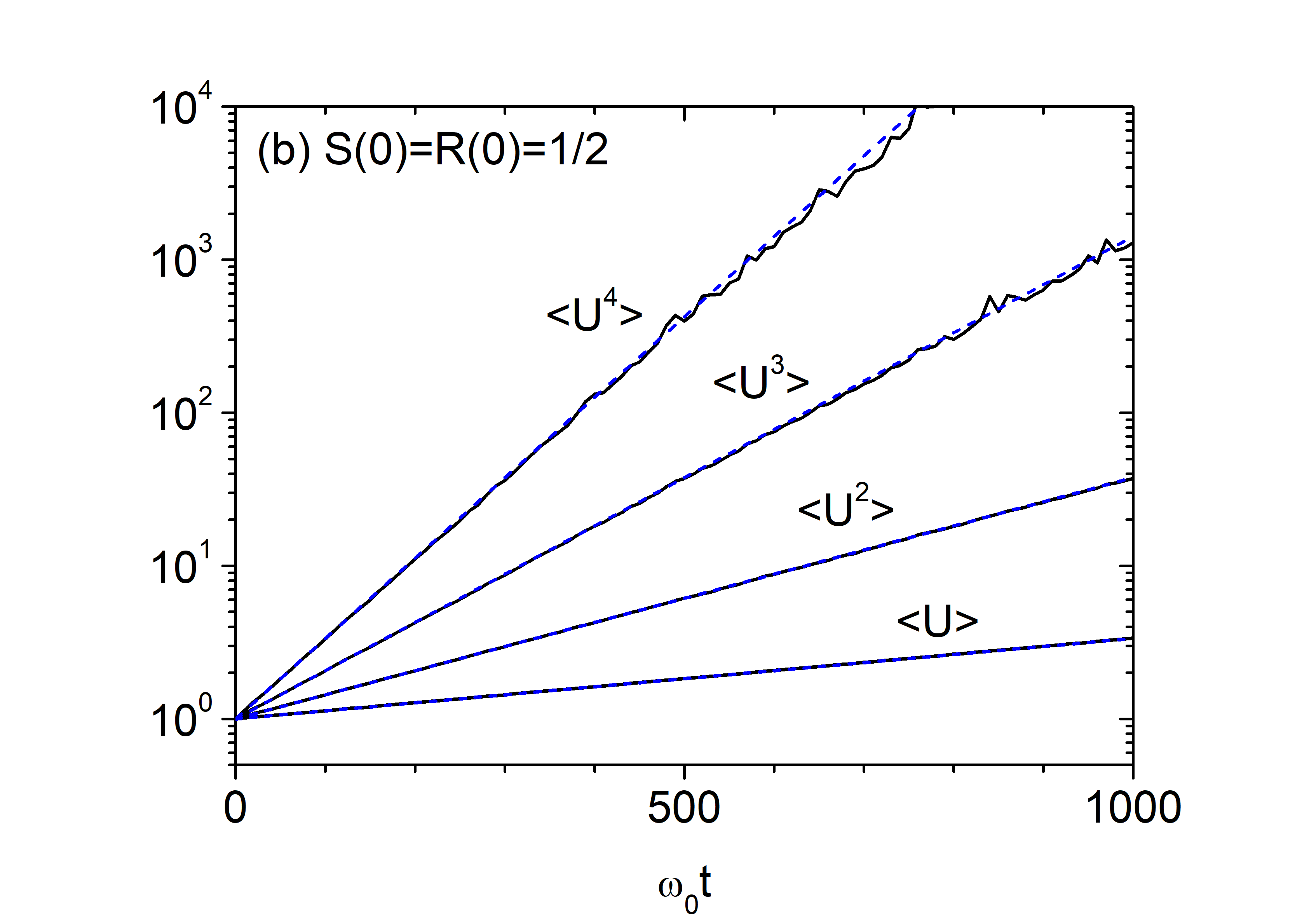}
    \caption{Temporal growth of energy moments under stepwise disorder (Model 2) with $\Lambda = 1/\omega_0$ and $a_0 = 0.1$.
Effective disorder strengths of $g \approx 0.00246$ (unidirectional input) and $g \approx 0.00242$ (symmetric bidirectional input) are obtained by fitting the exponential growth of $\langle U \rangle$.
(a) Numerical results (solid lines) for unidirectional input [$S(0) = 1$, $R(0) = 0$] agree well with the analytical prediction from Eq.~(\ref{eq:u1}) (dashed lines) at long times.
(b) For symmetric bidirectional input [$S(0) = R(0) = 1/2$], numerical results match the analytical expression from Eq.~(\ref{eq:u2}) at all times.}
\label{ff7}
  \end{figure}

   \begin{figure}
    \centering
    \includegraphics[width=\columnwidth]{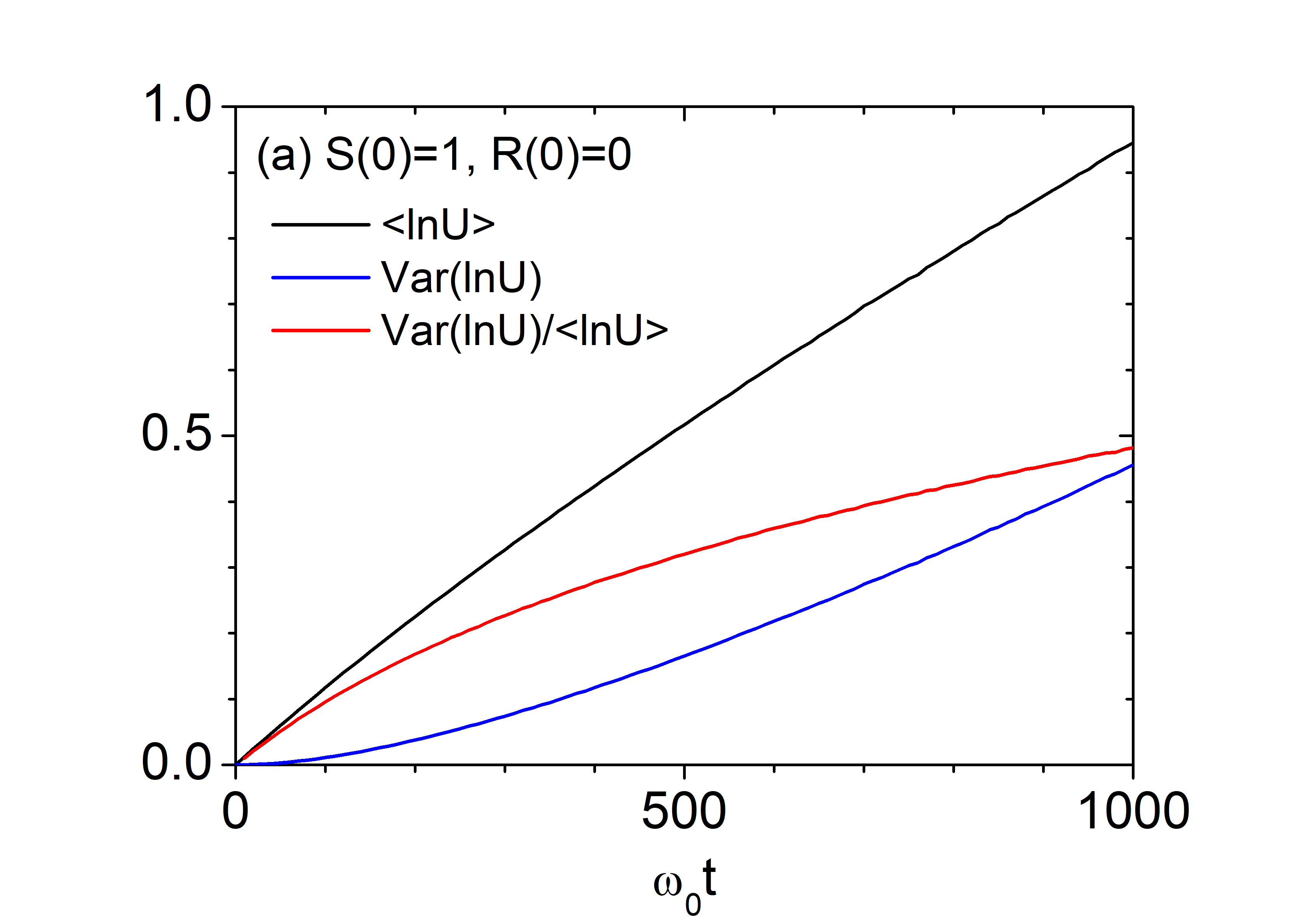}
   \includegraphics[width=\columnwidth]{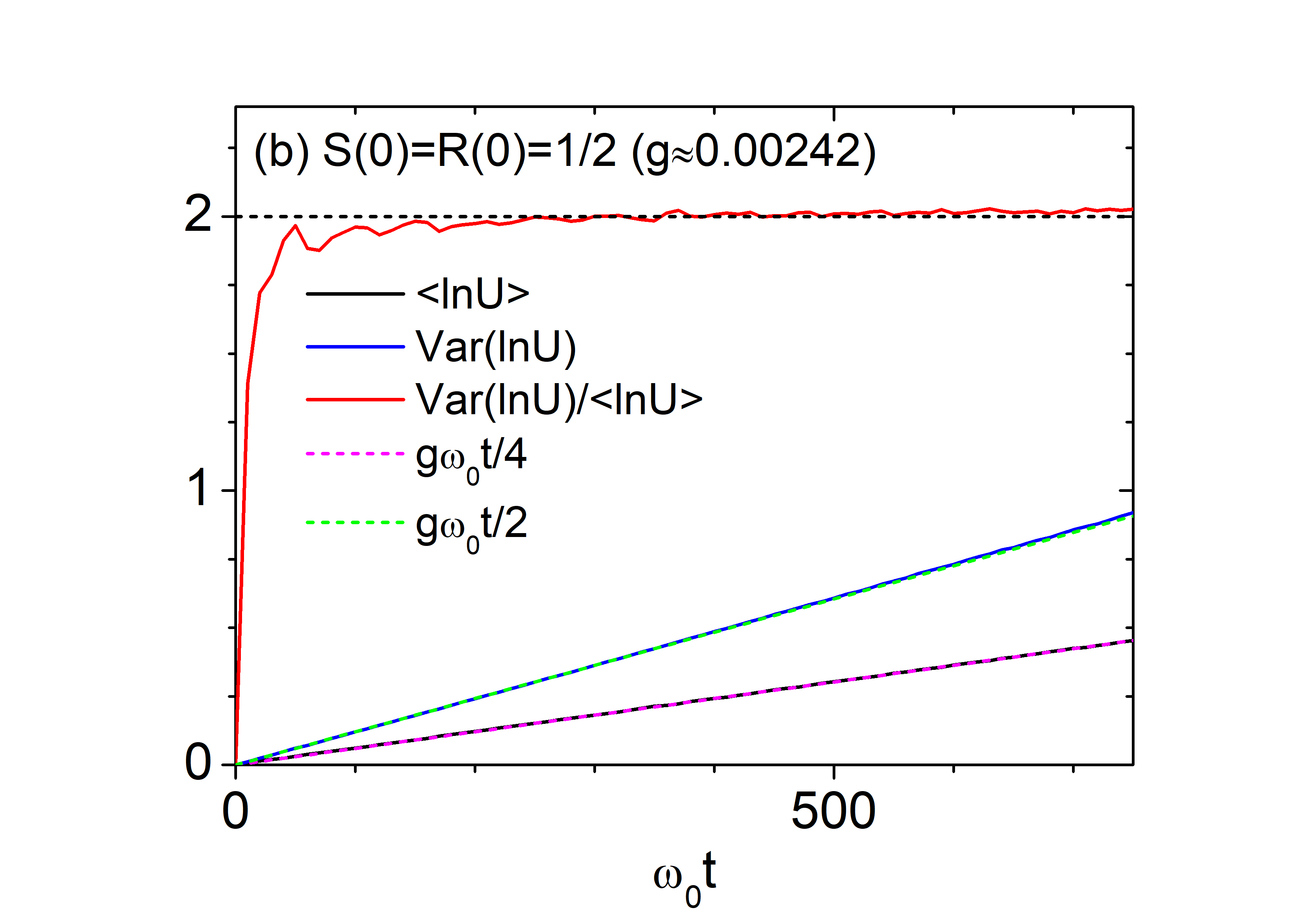}
    \caption{Temporal evolution of $\langle \ln U \rangle$, ${\rm Var}(\ln U)$, and their ratio under stepwise disorder (Model 2) with $\Lambda = 1/\omega_0$ and $a_0 = 0.1$.
Results are averaged over $10^6$ random configurations.
(a) Unidirectional input [$S(0) = 1$, $R(0) = 0$].
(b) Symmetric bidirectional input [$S(0) = R(0) = 1/2$] with effective disorder strength $g \approx 0.00242$, showing rapid convergence to the log-normal value ${\rm Var}(\ln U) / \langle \ln U \rangle = 2$.}
\label{ff8}
  \end{figure}

     \begin{figure}
    \centering
    \includegraphics[width=\columnwidth]{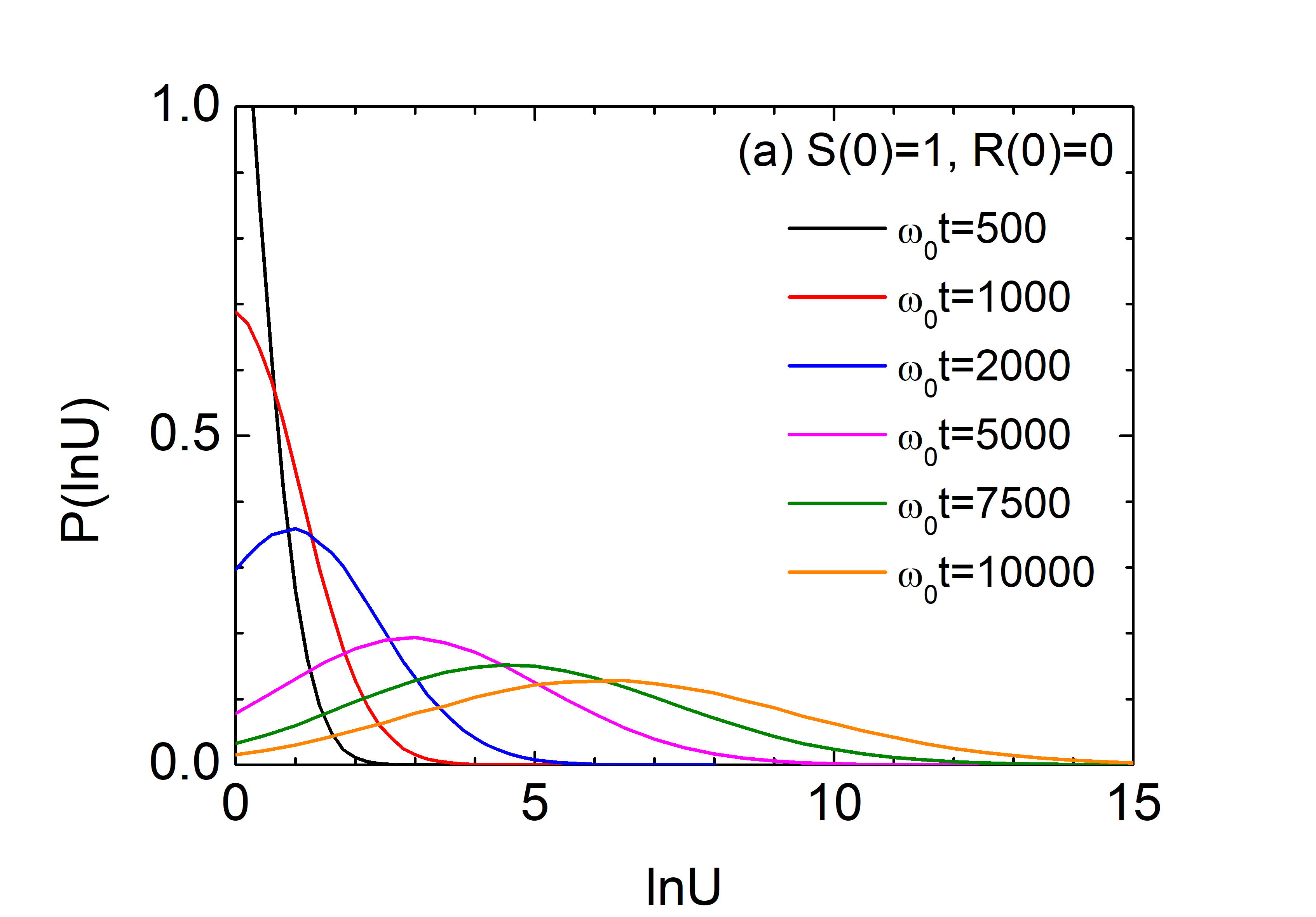}
    \includegraphics[width=\columnwidth]{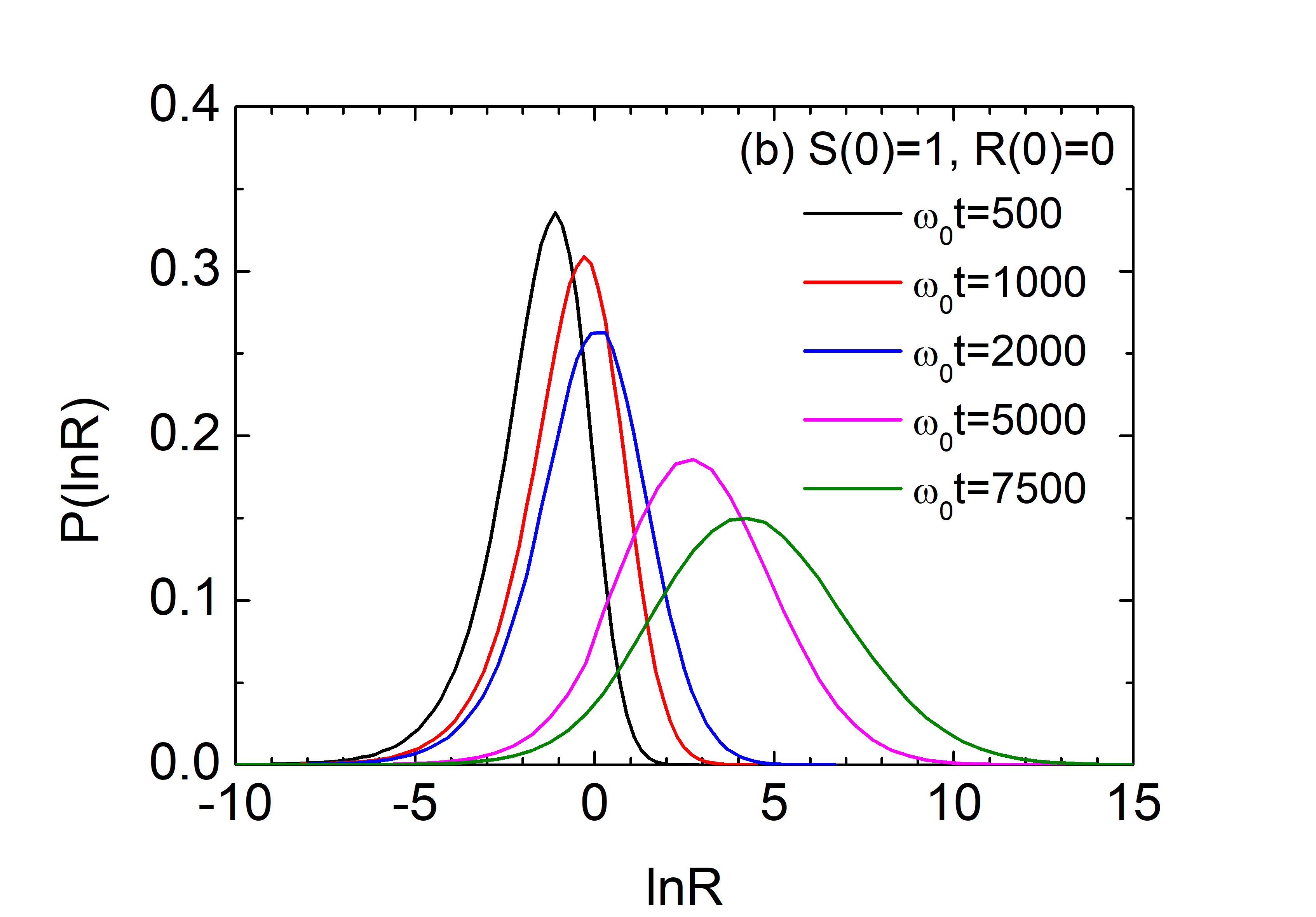}
    \includegraphics[width=\columnwidth]{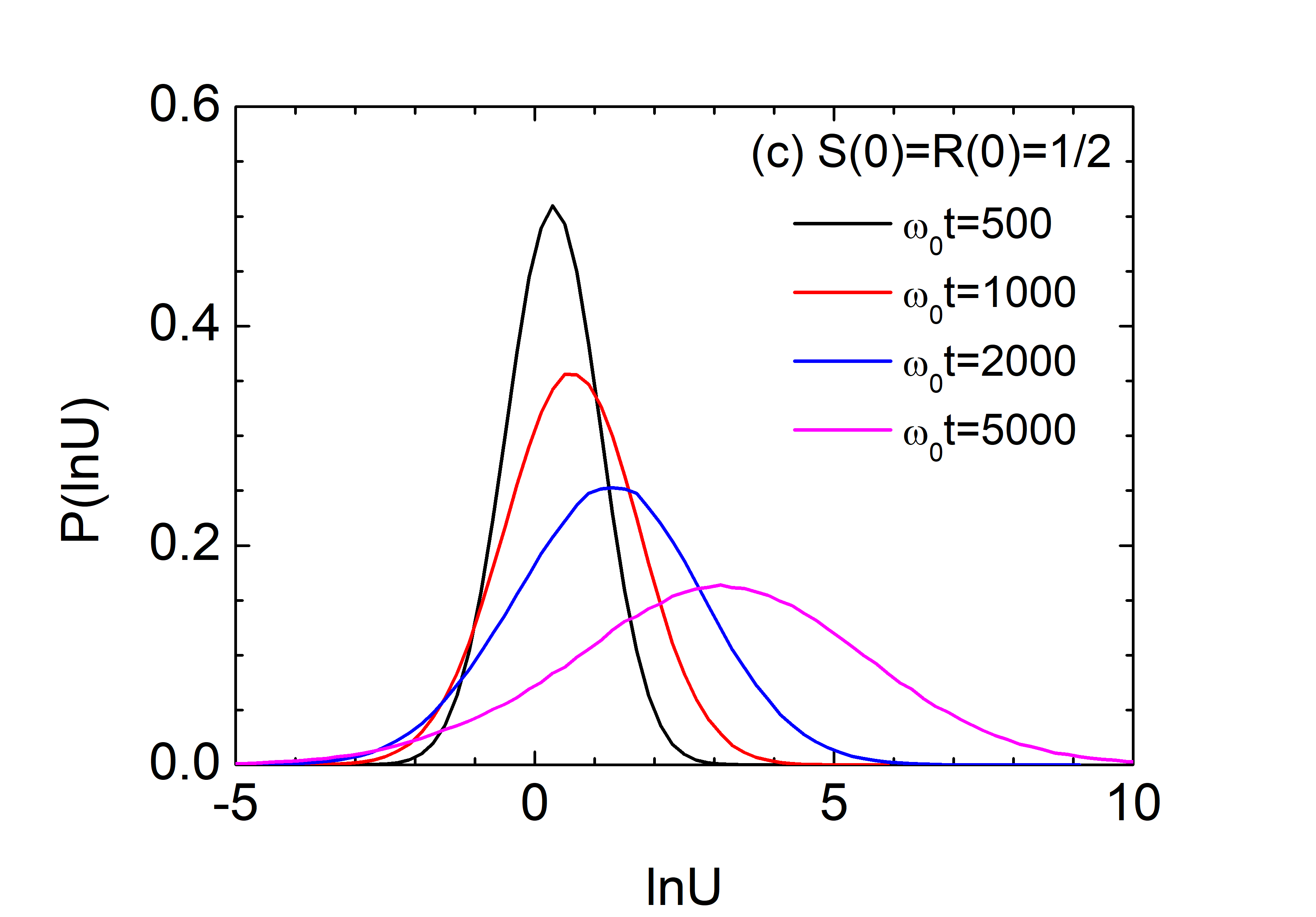}
    \caption{Evolution of probability density functions under stepwise disorder (Model 2) with $a_0 = 0.1$ and $\Lambda = 1/\omega_0$.
(a,b) Distributions of $\ln U$ and $\ln R$ for unidirectional input, obtained from histograms of $10^6$ samples at various times.
(c) Distribution of $\ln U$ for symmetric bidirectional input, showing Gaussian behavior consistent with log-normal statistics.}
\label{ff9}
  \end{figure}

In Fig.~\ref{ff3}, we plot the scaled energy moments
\begin{eqnarray}
U_n=\left\langle U^n\right\rangle \exp{\left[-n(n+1)\left(\frac{g\omega_0t}{4}\right)\right]},
\end{eqnarray}
for $n=1,2,3,4$.
In the long-time regime ($\omega_0t \gtrsim g^{-1}$), these values converge accurately to $n!/(2n-1)!!$ up to $\omega_0t\sim 10^6$. Beyond this point, deviations emerge due to the rapid growth of $\langle U^n \rangle$ and the accumulation of numerical errors.

To characterize the reflectance distribution more precisely, we evaluate the dimensionless quantity
 \begin{align}
  F_3=\frac{\left\langle R^3\right\rangle \left\langle R\right\rangle^3}{\left\langle R^2\right\rangle^3},
  \end{align}
which eliminates time-dependent scaling and isolates the distribution’s intrinsic form. The value of $F_3$ serves as a diagnostic: $F_3 = 5/9 \approx 0.556$ for a gamma distribution, 0.75 for a negative exponential, 1 for a log-normal, and 1.35 for the quasi-log-normal form given by Eq.~(\ref{eq:u1}). As shown in Fig.~\ref{ff4}, three distinct temporal regimes emerge: gamma-like for $\omega_0 t \lesssim 1$, negative exponential for $1 \lesssim \omega_0 t \lesssim g^{-1}$, and quasi-log-normal for $g^{-1} \lesssim \omega_0 t \lesssim 10^6$. Notably, $F_3$ never reaches 1, indicating that $R$ does not follow a true log-normal distribution. A similar trend is observed in the wave energy, with $\langle U^3 \rangle \langle U \rangle^3 / \langle U^2 \rangle^3 \approx 1.35$ at long times, consistent with quasi-log-normal statistics.
Since $g = g_0 \omega_0$, the crossover from negative exponential to quasi-log-normal behavior occurs at a time scale proportional to $g_0^{-1} \omega_0^{-2}$.

The statistical behavior is highly sensitive to the initial conditions.
When waves initially propagate in both directions, the moments are determined by the initial condition
$Z_{abcd}(0) = {w}^{a} ({w}^*)^b {v}^{c} ({v}^*)^d$, with ${w}$ and ${v}$ defined in Eq.~(\ref{eq:gic}). The resulting statistics depend on the relative magnitudes of ${w}$ and ${v}$.
Figure~\ref{ff5} shows the scaled moments $U_n$ ($n = 1$ to $5$) as functions of $R(0) = |{w}|^2$ at $\omega_0 t = 25000$, with $S(0) = |{v}|^2$ fixed at 1.
As $R(0)$ increases from 0 to 1—corresponding to symmetric bidirectional wave input—all $U_n$ increase and converge to 1.
In this limit, the energy distribution becomes exactly log-normal, with $U_n = 1$ for all $n$.

This behavior is confirmed in Fig.~\ref{ff6}, which shows the first four moments of $U$ for the symmetric case $R(0) = S(0) = 1/2$.
The numerical results exhibit excellent agreement with the analytical prediction given by Eq.~(\ref{eq:u2}) at all times, confirming that the energy follows a log-normal distribution throughout the evolution.

All numerical results in Figs.~\ref{ff3}–\ref{ff6} confirm that, in the long-time regime, the moments $\langle R^n\rangle$ and
$\langle U^n\rangle$ follow the exponential scaling of Eq.~(\ref{eq:uscal}). For $\langle U^n\rangle$, the prefactor is
$n!/(2n-1)!!$ for unidirectional input and unity for symmetric bidirectional input. For more general initial conditions, the prefactor is expected to vary systematically with the ratio
$w/v$. The observed trends—particularly the clear dependence on the initial condition in Fig.~\ref{ff5}—provide strong evidence that this behavior persists in the infinite-time limit.

At this stage, it is instructive to compare our findings with those of \cite{carm}, which studied wave propagation in a spatially uniform, time-varying medium in the weak-disorder regime using the transfer matrix method.
They reported that, in the time domain
$\tau_m\ll t\ll \tau_c$, the wave energy follows a negative exponential distribution, whereas for
$t\gg\tau_c$ it follows a log-normal distribution.
Our results for the intermediate‐time regime ($\tau_m \ll t \ll \tau_c$) are in excellent agreement with theirs. In our framework, the microscopic time $\tau_m$ corresponds to $1/\omega_0$, and the crossover time $\tau_c$ to $1/(g\omega_0)$.
Their Fokker–Planck equation [Eq.~(22)] contains the term
$z+z^2$ on the right-hand side.
The negative exponential result in \cite{carm} was obtained by neglecting the
$z^2$ term, while the log-normal result arose from neglecting the
$z$ term.
From their definitions,
$z$ corresponds to our
$R$ and
$z+1$ to our
$S$, so
$z+z^2$ maps directly to our
$RS$.
Neglecting the
$z^2$ term corresponds to the short-time regime, while neglecting the
$z$ term is equivalent to assuming
$R=S$, i.e., identical propagation in both directions.
Only in this latter limit does one recover a genuine log-normal distribution.
Retaining both $z$ and $z^2$ terms naturally leads to a distribution distinct from the log-normal form.

We also remark on the scaling characteristics of the underlying probability distribution. In the long-time regime, it is governed by two parameters:
$g\omega_0t/4$, which controls the exponential scaling of the moments, and the ratio
$w/v$, which sets the time-independent prefactor. For fixed initial conditions,
$w/v$ is constant, leaving
$g\omega_0t/4$ as the sole relevant parameter. In this qualified sense, the distribution exhibits single-parameter scaling \cite{bsha2,bsha1}, although, as in conventional spatial localization, this behavior is expected to break down under strong disorder.

We further performed simulations using Model 2, where the random component $\delta\epsilon$ remains constant over each time interval of duration $\Lambda = 1/\omega_0$ and is abruptly updated at the end of each interval. Each value is independently sampled from a uniform distribution over $[-a_0, a_0]$, with $a_0 = 0.1$. Since $\overline\epsilon=1$, $\epsilon(t)$ remains strictly positive within the range $[0.9,1.1]$. Equation~(\ref{eq:iie}) was solved for $10^6$ independent realizations, and the results were ensemble averaged over the total duration $T$.

The average energy $\langle U \rangle$ grows exponentially at all times, regardless of the initial condition. Fitting this growth yields effective disorder strengths of $g \approx 0.00246$ for unidirectional input and $g \approx 0.00242$ for symmetric bidirectional input. Using these fitted values, we compared the numerical results from Model 2 with the analytical predictions of Eqs.~(\ref{eq:u1}) and (\ref{eq:u2}). As shown in Fig.~\ref{ff7}(a), the energy moments for unidirectional input closely follow Eq.~(\ref{eq:u1}) in the long-time regime. For symmetric bidirectional input, the moments agree with the log-normal form of Eq.~(\ref{eq:u2}) at all times [Fig.~\ref{ff7}(b)]. These results confirm that the statistical behavior is robust and essentially independent of the disorder model, provided the disorder is weak and short-range correlated.

For Model 2 with $\Lambda = 1/\omega_0$ and $a_0 = 0.1$, we also computed the mean and variance of $\ln U$.
Figures~\ref{ff8}(a) and \ref{ff8}(b) show $\langle \ln U \rangle$, ${\rm Var}(\ln U)$, and their ratio for unidirectional and symmetric bidirectional inputs.
For symmetric input, the results agree with the log-normal predictions: $\langle \ln U \rangle = g\omega_0 t / 4$, ${\rm Var}(\ln U) = g\omega_0 t / 2$, yielding a ratio of 2.
In contrast, unidirectional input exhibits persistent deviations, with the ratio remaining well below 2 even at long times—indicating a clear departure from log-normal behavior.

Finally, we computed the probability distributions of $\ln U$ and $\ln R$ using Model 2 under stepwise disorder with $\Lambda = 1/\omega_0$ and $a_0 = 0.1$, based on $10^6$ independent simulations.
Figures~\ref{ff9}(a) and \ref{ff9}(b) show histogram-based probability density functions of $\ln U$ and $\ln R$ at various times for unidirectional input. Since $U \geq 1$, $\ln U$ is always positive, while $\ln R$ spans the entire real axis. The distribution of $\ln U$ undergoes a clear crossover from an exponential-like form to a quasi-log-normal shape, while the distribution of $\ln R$ evolves from being highly skewed to increasingly symmetric over time. For symmetric bidirectional input [Fig.~\ref{ff9}(c)], the distribution of $\ln U$ remains Gaussian at all times, consistent with genuine log-normal behavior.

\section{Discussion}

Our study shows that wave propagation in randomly time-varying media exhibits rich statistical behavior that differs fundamentally from that in spatially disordered systems. While earlier work characterized temporal Anderson localization in terms of negative exponential and log-normal statistics \cite{carm}, we demonstrate that this picture is incomplete. In particular, the energy distribution is strongly influenced by the initial wave configuration—an effect largely overlooked in previous analyses.

For unidirectional input, the statistics evolve through three distinct regimes: a gamma distribution at early times, negative exponential behavior in the intermediate regime ($1 \lesssim \omega_0 t \lesssim g^{-1}$), and a quasi-log-normal distribution at long times. The latter deviates from a true log-normal form due to a distinct prefactor in the moments, originating from temporal interference and amplitude fluctuations.
In contrast, symmetric bidirectional input yields genuine log-normal statistics across all time scales, as confirmed by analytical predictions, numerical simulations, and agreement in the scaled moments $U_n$ and the ratio ${\rm Var}(\ln U)/\langle \ln U \rangle$.

Momentum conservation plays a central role: although the total energy increases under temporal driving, the difference between reflected and transmitted energies remains constant, directly linking the initial wave symmetry to the long-term statistical behavior.
Our results are consistent across two distinct disorder models—delta-correlated Gaussian noise and stepwise uniform fluctuations—indicating that the observed statistical behavior reflects universal properties of temporally disordered systems, rather than model-specific artifacts, in the spirit of renormalization group theory. We therefore expect similar statistics to arise in any short-range-correlated random model within the weak-disorder regime, provided finite-size effects are negligible.

These findings have direct implications for dynamic wave control and advanced device design. By precisely tailoring the initial wave symmetry and key parameters of temporal disorder—such as its duration, strength, and correlation length—it is possible to engineer targeted energy distributions \cite{jkim}. Such control could enable applications in photonic switching, robust energy confinement, and temporally programmable media.

A principal limitation of this study is the assumption of an infinitely long medium along the propagation direction, which effectively neglects reflections from spatial boundaries. This assumption holds when the medium length greatly exceeds the wavelength and the observation time is much shorter than the wave traversal time across the medium.

Finite-size effects, particularly when the medium length is comparable to the wavelength, have been investigated in previous studies of wave propagation in time-varying media for both periodic \cite{zur,vald} and random \cite{bo} temporal variations. In the random case, waves confined in a cavity of similar size to the wavelength, under random permittivity variations, have been shown to exhibit a nontrivial L\'evy-type distribution \cite{bo}. A promising direction for future research is to examine finite-size effects in regimes where the medium length exceeds the wavelength and boundary scattering plays only a perturbative role.

In the strong-disorder regime, qualitatively different statistical behavior is expected. Future work will address this case using a model with bounded dichotomous disorder in $\delta\epsilon$, analyzed via the Shapiro–Loginov formula of differentiation \cite{shalog}. Other promising avenues include extending the formalism to systems with long-range–correlated disorder.

Experimental implementations using metamaterials, time-modulated dielectrics, or plasmas could provide valuable platforms to test and exploit the statistical regimes identified in this work.

\begin{acknowledgments}
This work was supported by the National Research Foundation of Korea (https://ror.org/013aysd81) grant funded by the Korean Government (RS-2025-16071339). It was also supported by the Basic Science Research Program through the National Research Foundation of Korea, funded by the Ministry of Education (RS-2021-NR060141).
\end{acknowledgments}

\section*{Appendix: Derivation of analytical expressions for reflectance moments in the short- and intermediate-time regimes}
\renewcommand{\theequation}{A\arabic{equation}}
\setcounter{equation}{0}

In this appendix, we present analytical derivations of the reflectance moments in the short‐ and intermediate‐time regimes, corresponding to Eqs.~(\ref{eq:short}) and (\ref{eq:exp}).
Both results are obtained via a perturbative expansion in the small duration $T$.
The derivation of Eq.~(\ref{eq:exp}) employs the RPA, whereas Eq.~(\ref{eq:short}) is derived without invoking the RPA.

When the duration $T$ is small, we expand the moment $Z_{abcd}$, where $a$, $b$, $c$, and $d$ are nonnegative integers, as a power series in $T$:
\begin{equation}
Z_{abcd}=Z_{abcd}^{(0)}+Z_{abcd}^{(1)}T+Z_{abcd}^{(2)}T^2+\cdots,
\label{eq:zps}
\end{equation}
where $Z_{abcd}^{(j)}$ is the coefficient of the \textit{j}-th order term. From the initial condition at $T = 0$, we obtain
\begin{equation}
Z_{abcd}^{(0)}=\begin{cases}
                1, & \mbox{if } a=b=0 \\
                0, & \mbox{otherwise}
              \end{cases}.
\label{eq:zps1}
\end{equation}
Substituting Eq.~(\ref{eq:zps}) into the governing equation [Eq.~(\ref{eq:z})] and applying the perturbation method shows that, for $Z_{nn00}$, only the terms proportional to $C_7$, $C_{11}$, and $C_{13}$ in Eq.~(\ref{eq:z}) contribute. This yields the recursion relation
\begin{align}
Z_{abcd}^{(j)}=&\frac{\eta}{j}\bigg[ab Z_{a-1,b-1,c+1,d+1}^{(j-1)}-\frac{a(a-1)}{2}Z_{a-2,b,c+2,d}^{(j-1)}\nonumber\\&
~~~~-\frac{b(b-1)}{2}Z_{a,b-2,c,d+2}^{(j-1)}\bigg],
\label{eq:recursion}
\end{align}
where
$\eta=g\omega_0/4$.
All coefficients are real, since no imaginary terms appear.

From Eqs.~(\ref{eq:zps1}) and (\ref{eq:recursion}), the first‐order term is
\begin{align}
Z_{1100}^{(1)} &= \eta Z_{0011}^{(0)} = \frac{g\omega_0}{4},
\end{align}
so that
\begin{align}
\langle R \rangle &\approx \frac{g\omega_0 T}{4}.
\end{align}
For $n \ge 2$, $Z_{nn00}^{(j)}$ vanishes for $j < n$, so the leading term is $\mathcal{O}(T^n)$.
For $n=2$ the second‐order term is
\begin{equation}
Z_{2200}^{(2)} = \frac{\eta}{2} \left( 4Z_{1111}^{(1)} - Z_{0220}^{(1)} - Z_{2002}^{(1)} \right),
\end{equation}
with $Z_{1111}^{(1)} = \eta$ and $Z_{0220}^{(1)} = Z_{2002}^{(1)} = -\eta$. Substituting these values gives $Z_{2200}^{(2)} = 3\eta^2$, and thus
\begin{align}
\langle R^2 \rangle &\approx 3\eta^2 T^2 = 3!! \left( \frac{g\omega_0 T}{4} \right)^2.
\end{align}
For $n=3$:
\begin{equation}
Z_{3300}^{(3)}=\frac{\eta}{3}\left(9Z_{2211}^{(2)}
-3Z_{1320}^{(2)}-3Z_{3102}^{(2)}\right),
\end{equation}
with $Z_{1320}^{(2)}=Z_{3102}^{(2)}$ and
\begin{align}
Z_{2211}^{(2)}&=\frac{\eta}{2}\left(4Z_{1122}^{(1)}
-Z_{0231}^{(1)}-Z_{2013}^{(1)}\right)=3\eta^2,\nonumber\\
Z_{1320}^{(2)}&=\frac{\eta}{2}\left(3Z_{0231}^{(1)}
-3Z_{1122}^{(1)}\right)=-3\eta^2,
\end{align}
leading to $Z_{3300}^{(3)}=15\eta^3$ and
\begin{align}
\langle R^3\rangle&\approx 15\eta^3 T^3=5!!\left(\frac{g\omega_0T}{4}\right)^3.
\end{align}
For $n=4$:
\begin{equation}
Z_{4400}^{(4)}=\frac{\eta}{4}\left(16Z_{3311}^{(3)}
-6Z_{2420}^{(3)}-6Z_{4202}^{(3)}\right),
\end{equation}
where $Z_{2420}^{(3)}=Z_{4202}^{(3)}$.
Using Eq.~(\ref{eq:recursion}) recursively, we obtain
\begin{align}
&Z_{3311}^{(3)}=\frac{\eta}{3}\left(9Z_{2222}^{(2)}
-3Z_{1331}^{(2)}-3Z_{3113}^{(2)}\right)\nonumber\\
&~~~=\frac{\eta^2}{6}\left[9\left(4Z_{1133}^{(1)}-2Z_{0242}^{(1)}\right)
-6\left(3Z_{0242}^{(1)}-3Z_{1133}^{(1)}\right)\right]\nonumber\\
&~~~=15\eta^3,\nonumber\\
&Z_{2420}^{(3)}=\frac{\eta}{3}\left(8Z_{1331}^{(2)}-Z_{0440}^{(2)}
-6Z_{2222}^{(2)}\right)\nonumber\\
&~~~=\frac{\eta^2}{6}\left[8\left(3Z_{0242}^{(1)}-3Z_{1133}^{(1)}\right)+6Z_{0242}^{(1)}\right.\nonumber\\
&~~~~~~~~~~~~\left.-6\left(4Z_{1133}^{(1)}-2Z_{0242}^{(1)}\right)\right]\nonumber\\
&~~~=-15\eta^3,
\end{align}
which yields $Z_{4400}^{(4)}=105\eta^4$ and
\begin{align}
\langle R^4\rangle&\approx 105\eta^4T^4=7!!\left(\frac{g\omega_0T}{4}\right)^4.
\end{align}
Repeating this procedure for arbitrary $n$ yields the general short‐time result [Eq.~(\ref{eq:short})]:
\begin{align}
\langle R^n\rangle&\approx (2n-1)!!\left(\frac{g\omega_0T}{4}\right)^n.
\end{align}

Under the RPA, the recursion relation takes a particularly simple form. Substituting Eq.~(\ref{eq:rpa}) into Eq.~(\ref{eq:recursion}) gives
\begin{align}
Z_{aacc}^{(j)}=\frac{a^2 \eta}{j}Z_{a-1,a-1,c+1,c+1}^{(j-1)}.
\label{eq:recursion2}
\end{align}
Iterating this relation yields
\begin{align}
Z_{nn00}^{(n)} &= n\eta\, Z_{n-1,n-1,1,1}^{(n-1)} \nonumber\\
               &= n\eta\, (n-1)\eta\, Z_{n-2,n-2,2,2}^{(n-2)} \nonumber\\
               &~~\vdots \nonumber\\
               &= n! \eta^n Z_{00nn}^{(0)} \nonumber\\
               &= n! \eta^n,
\end{align}
leading to Eq.~(\ref{eq:exp}):
\begin{align}
\langle R^n\rangle&\approx n!\eta^nT^n=n!\left(\frac{g\omega_0T}{4}\right)^n.
\end{align}

\end{document}